\documentclass[amssymb,amsmath,aps,prb,twocolumn,superscriptaddress,a4paper]{revtex4}

\usepackage[dvips]{graphicx}
\usepackage[english]{babel}
\usepackage[usenames]{color}
\usepackage[normalem]{ulem}
\usepackage{amsfonts}
\usepackage{amssymb}
\usepackage{amsmath}

\begin{document}

\title{Magnon softening in a ferromagnetic monolayer: a first-principles spin dynamics study}

\author{Anders Bergman}
\email{anders.bergman@fysik.uu.se}
\affiliation{Department of Physics and Astronomy, Uppsala University, Box 516, 751~20 Uppsala, Sweden}
\affiliation{Department of Materials Science and Engineering, Royal Institute of Technology, Brinellv\"{a}gen 23, 100~44 Stockholm, Sweden}
\author{Andrea Taroni}
\affiliation{Department of Physics and Astronomy, Uppsala University, Box 516, 751~20 Uppsala, Sweden}
\author{Lars Bergqvist}
\affiliation{Department of Physics and Astronomy, Uppsala University, Box 516, 751~20 Uppsala, Sweden}
\author{Johan Hellsvik}
\affiliation{Department of Physics and Astronomy, Uppsala University, Box 516, 751~20 Uppsala, Sweden}
\author{Bj\"orgvin Hj\"orvarsson}
\affiliation{Department of Physics and Astronomy, Uppsala University, Box 516, 751~20 Uppsala, Sweden}
\author{Olle Eriksson}
\affiliation{Department of Physics and Astronomy, Uppsala University, Box 516, 751~20 Uppsala, Sweden}


\begin{abstract}
We study the Fe/W(110) monolayer system through a combination of first principles calculations and atomistic spin dynamics simulations. We focus on the dispersion of the spin waves parallel to the [001] direction. Our results compare favorably with the experimental data of Prokop \emph{et al.} [\emph{Phys. Rev. Lett.} \textbf{102}, 177206], and correctly capture a drastic softening of the magnon spectrum, with respect to bulk bcc Fe. The suggested shortcoming of the itinerant electron model, in particular that given by density functional theory, is refuted. We also demonstrate that finite temperature effects are significant, and that atomistic spin dynamics simulations represent a powerful tool with which to include these.
\end{abstract}

\keywords{Atomistic spin dynamics; monolayer magnetism; ultrathin magnets; magnons; spin waves; spin Hamiltonians; iron; tungsten}
\pacs{75.30.Ds,75.50.Bb,75.70.Ak,75.40.Mg}
\maketitle

Recent progress in experimental techniques have allowed the first observation of the magnon dispersion spectrum of a single ferromagnetic monolayer~\cite{Prokop2009}. The measurements, which were performed on an atomic monolayer (ML) of Fe on W(110) using spin-polarized electron energy loss spectroscopy (SPEELS), revealed magnon energies that are much smaller compared to the bulk and surface Fe(110) excitations. This strong magnon softening is in contradiction with theoretical predictions based on an itinerant electron model at $T=0$~K~\cite{Muniz2002,Costa2008}. This discrepancy raises the possibility that a ML of Fe on W(110) may not be a simple itinerant ferromagnet, as generally assumed, and indicates a shortcoming in the theoretical understanding of low-dimensional systems. In fact, from their experimental results the authors of Ref.~\onlinecite{Prokop2009} suggested this possibility. Furthermore, a recent theoretical study by Grechnev \emph{et al.} has suggested that effects of electron correlations might be more important for surfaces~\cite{Grechnev2007}. Given the fundamental nature of spin wave excitations, and their role in physical processes such as fast magnetization reversal~\cite{Tudosa2004} and current induced magnetic switching~\cite{Fuchs2004,Sankey2008,Kubota2008}, it has become both timely and important to address these questions.
 
\par
Low-dimensional magnetic structures lack inversion symmetry. Consequently, significant anisotropic contributions to the magnetic ordering can arise, in the form of Dzyaloshinskii-Moriya interactions (DMI)~\cite{Dzyaloshinskii1957,Moriya1960} and magnetic anisotropy energies. Both of these stem from relativistic spin orbit coupling, and may be energetically strong enough to compete with isotropic exchange interactions, leading to complex magnetic ground states in systems such as Mn on W(001)~\cite{Bode2007} and Fe on W(110)~\cite{Heide2008,Udvardi2009}. Furthermore, the interaction between the Fe monolayer and the W substrate is known to be significant, and in the W(001) case this can even lead to antiferromagnetic ordering~\cite{Wu1992,Kubetzka2005,Ferriani2005,Sandratskii2006}. An additional challenge when modeling spin waves is posed by temperature: the experimental data in Ref.~\onlinecite{Prokop2009} were taken at 120~K, which corresponds to a value for $T/T_{\mathrm{c}}=0.5$, where $T_{\mathrm{c}}$ is the Curie temperature. In this regime, finite temperature effects, which are normally excluded from density functional theory, can be expected to play an important role. 

\par
We demonstrate that atomistic spin dynamics (ASD) simulations based on first principles theory provide a powerful tool for studying magnons at finite temperature. Using this approach, we are able to account for most of the experimentally observed magnon softening in one ML Fe on W(110)~\cite{Prokop2009}. We mainly attribute the softening to hybridization with the underlying W substrate combined with finite temperature effects. We show that the itinerant electron model of surface and thin film magnetism, as given by density functional theory, does not have to be abandoned, provided dynamical aspects and finite temperature effects are considered.

\par
We performed atomistic spin dynamics simulations~\cite{Antropov1996} using the UppASD package~\cite{Skubic2008}. In the simulations, the isotropic exchange interactions are treated by introducing the classical Hamiltonian
\begin{equation}
  \mathcal{H} = -\frac{1}{2}\sum_{i\neq j} J_{ij} \mathbf{m}_i \cdot \mathbf{m}_j,
\label{eqn:hh}
\end{equation}
\noindent where $i$ and $j$ are atomic indices, $\mathbf{m}_i$ is a classical atomic moment and $J_{ij}$ is the strength of the exchange interaction. Relativistic effects are taken into account by adding one or both terms from the equation
\begin{equation}
 \mathcal{H}_{\mathrm{SO}} = K \sum_{i}\left(\mathbf{m}_i \cdot \mathbf{e}_K\right) ^{2} + \sum_{i,j} \mathbf{D}_{ij} \left(\mathbf{m}_i \times \mathbf{m}_j\right),
 \label{eqn:so}
\end{equation}
\noindent where $K$ is the strength of the anisotropy field along the direction of $\mathbf{e}_K$, and $\mathbf{D}_{ij}$ is the Dzyaloshiskii-Moriya vector~\cite{Dzyaloshinskii1957,Moriya1960}. From this Hamiltonian, the effective interaction field experienced by each atomic moment $\mathbf{m}_i$ is calculated as $\mathbf{B}_i = - \frac{\partial\mathcal{H}}{\partial\mathbf{m}_i}$. The temporal evolution of the atomic spins at finite temperature is modeled by Langevin dynamics, through coupled stochastic differential equations of the Landau-Lifshitz form, 

\[
\frac{\partial\mathbf{m}_i}{\partial t} = - \gamma\mathbf{m}_i \times \left[\mathbf{B}_i + \mathbf{b}_i(t)\right] - \gamma\frac{\alpha}{m} \mathbf{m}_i \times \left\{ \mathbf{m}_i \times\left[\mathbf{B}_i + \mathbf{b}_i(t)\right] \right\}.
\]

\noindent In this expression, $\gamma$ is the the electron gyromagnetic ratio, and $\mathbf{b}_i$ is a stochastic magnetic field with a Gaussian distribution, the magnitude of which is related to the temperature and the phenomenological damping parameter $\alpha$, which eventually brings the system to thermal equilibrium.

\par
We have calculated the exchange parameters, $J_{ij}$, from first principles calculations. A supercell geometry containing seven W layers, terminated with a single Fe layer on each side and surrounded with vacuum, was used. In all calculations, the in-plane lattice constant ($a_{lat}$) was fixed to the experimental value of bcc W (3.165 \AA). The geometrical ground state was obtained by performing a structural relaxation of the slab using the projector augmented wave (PAW) method~\cite{Kresse1999}, as implemented in the Vienna ab-initio simulations package (VASP)~\cite{Kresse1996}. We found an equilibrium geometry in which there was a significant relaxation of the Fe-W distance (-13\%), in agreement with previous reports~\cite{Albrecht1991,Quian2003}.

\par
The relaxed structure was used as the input for the calculation of the $J_{ij}$ values, using both the Liechtenstein-Katsnelson-Gubanov method (LKGM)~\cite{Liechtenstein1984,Liechtenstein1986}, and the ``frozen magnon" approximation~\cite{Halilov1997,Halilov1998}. In the first case, the exchange interaction parameters are obtained from small angle perturbations from the ground state, using the magnetic force theorem~\cite{Andersen1980}. In the second case, they are obtained by an inverse Fourier transform of the total energy $E(\mathbf{q})$ for a large number of spin-spirals with wave vector $\mathbf{q}$. For a ferromagnet, this latter method also has the advantage of directly providing the adiabatic magnon spectra at 0~K~\cite{Halilov1998}, therefore providing a benchmark with which to compare the results of the ASD simulations, and to verify that the spin dynamics Hamiltonian is correct.

\par
In order to study the effect of finite temperature on the exchange interactions, we performed the LKGM calculations both for a ferromagnetic configuration, corresponding to the magnetic order at 0~K, and in Disordered Local Moments (DLM) states~\cite{Oguchi1983,Pindor1983}, which give a better description the magnetic order at high temperatures. The LKGM calculations were performed using a Korringa-Kohn-Rostoker (KKR) method, while a Linear Muffin Tin Orbital method (LMTO)~\cite{Andersen1975} was used for the frozen magnon calculations. Both KKR and LMTO calculations were performed within the atomic sphere approximation, using the Local Spin Density Approximation. The anisotropy constant we used in the spin dynamics simulations was taken from experimental data, and corresponds to $2K_{\mathrm{eff}}S = 4.6$ meV~\cite{Prokop2009,Pratzer2001}.

\par
The LMTO-based spin-spiral calculations suggest that, rather than a ferromagnetic solution, the global energy minimum is given by a spin-spiral with $\mathbf{q}=[0~0~0.1]$ \AA$^{-1}$, which is in agreement with recent full-potential plane wave calculations~\cite{Nakamura2007}. This spin-spiral state was also found when evaluating the exchange parameters obtained from the LKGM calculations for the ferromagnetic configuration, but with an energy difference on the same scale as the maximal energy resolution of the method. The exchange parameters calculated from DLM configurations correspond to a ferromagnetic solution. Although a spin-spiral energy minimum appears to contradict the widely held view that 1 ML of Fe on W(110) is ferromagnetic~\cite{Prokop2009,Udvardi2009,Costa2008}, the difference in energy between the ferromagnetic and spin-spiral state is less than 1~meV. Thus, the ferromagnetic state is stabilized by the magnetic crystalline anisotropy. This is confirmed in our spin dynamics simulations, by adding the first term of Eq. (\ref{eqn:so}) to the Heisenberg Hamiltonian (\ref{eqn:hh}).

\begin{figure}
  \includegraphics[width=8.3cm]{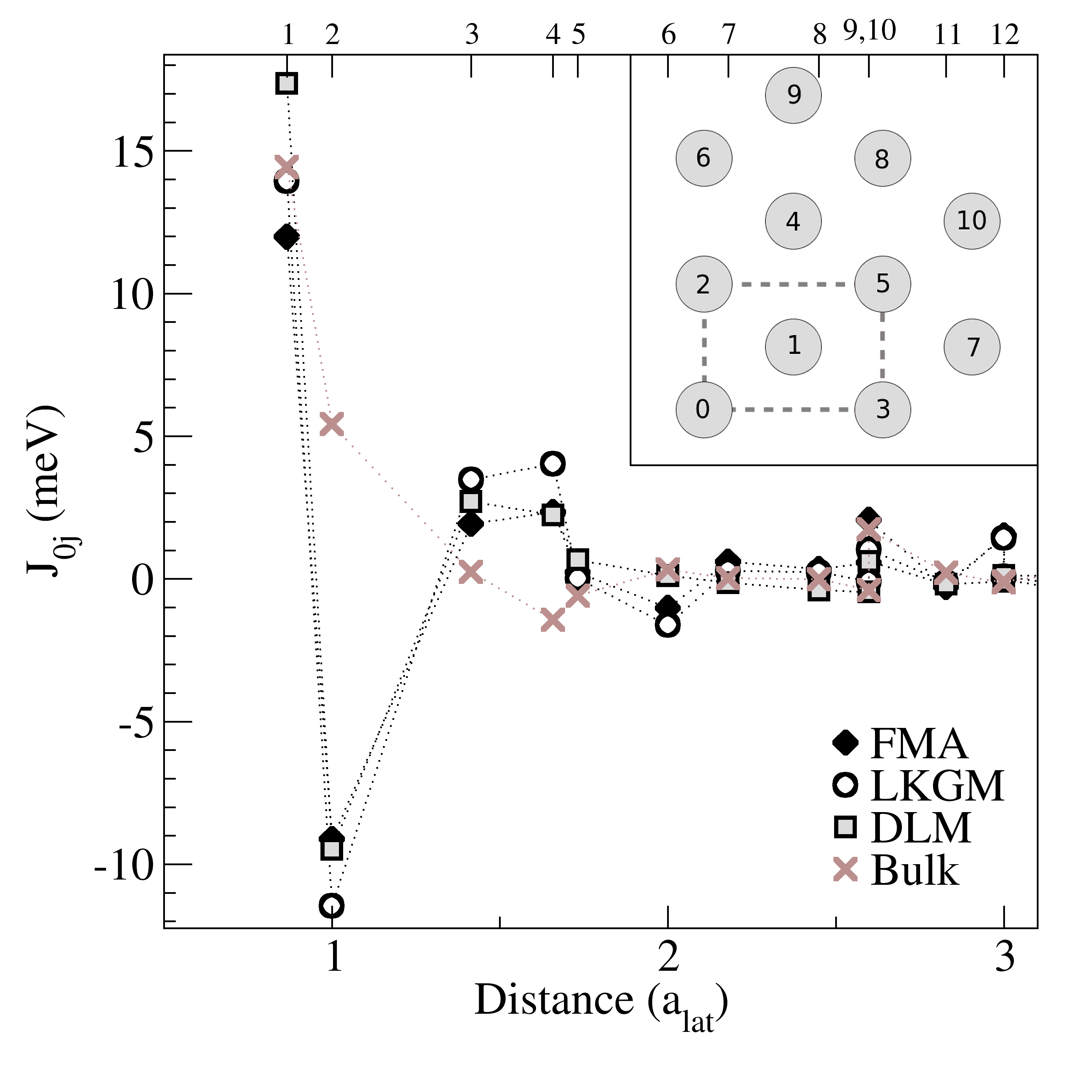}
  \caption{\label{fig:jijs}(Color online) Calculated exchange interaction parameters $J_{ij}$ for a ML Fe on W(110) as a function of distance (in lattice constant $a_{lat}$). The $J_{ij}$s labelled FMA were obtained using the ``frozen magnon" approximation while the other curves was calculated using the LKGM method for a ferromagnetic solution (LKGM), and disordered local moment state (DLM). Also shown are the calculated exchange interaction parameters for bulk bcc Fe. The inset shows the geometry of the ML and position of neighbour $j$ relative to site $0$. The $J_{ij}$ values in the figure have been scaled with the square of the magnetic moment to follow the standard convention of Ref.~\onlinecite{Liechtenstein1984}.}
\end{figure}

\par
The interatomic exchange parameters calculated both from LKGM and frozen magnon approaches are shown in Fig.~\ref{fig:jijs}. Although the nearest-neighbor interaction dominates, longer-range interactions can also be seen to be significant. The obtained values display the same trend as recent calculations up to the fourth shell of neighbors~\cite{Udvardi2009}, but we note that our value for next-nearest exchange interaction is larger in magnitude. Furthermore, we have also calculated $J_{ij}$s for the unrelaxed surface structure (not shown). In this case, the nearest-neighbour interaction strength increases by $40\%$, whereas the magnitude of the next-nearest interaction decreases by $50\%$, relative to the relaxed structure. Calculating the spin-wave spectra from the unrelaxed set of exchange parameters results in a stiffer magnon spectrum compared to the relaxed system,  a result that underlines the importance of hybridization effects between Fe atoms and the W substrate. We emphasize that in order to correctly reproduce the $T=0$~K magnon spectra from the frozen magnon calculation with the spin dynamics simulations, at least 20 shells of neighbors had to be included in the Heisenberg Hamiltonian. With the LKGM approach, the convergence was even more sensitive, with up to 80 shells required. 

\par
Using the first principles exchange interactions in the atomistic spin dynamics Hamiltonian allows us to address the dynamical properties of spin systems at finite temperatures~\cite{Skubic2008,Tao2005,Chen1994}. Of particular interest to us is the dynamical structure factor $S^k(\mathbf{q},\omega)$, which is the quantity probed in neutron scattering experiments of bulk systems~\cite{Lovesey1984}, and can analogously be applied to SPEELS measurements. The dynamical structure factor is readily obtained by a Fourier transform of the space- and time-displaced correlation function, 

\[
C^k (\mathbf{r}-\mathbf{r'},t) = \langle m_{\mathbf{r}}^k(t) m_{\mathbf{r'}}^k(0) \rangle - \langle m_{\mathbf{r}}^k(t) \rangle \langle m_{\mathbf{r'}}^k(0) \rangle,
\]

\noindent where the angular brackets signify an ensemble average, and $k$ the cartesian component. 

\begin{figure}
  \includegraphics[width=8.4cm]{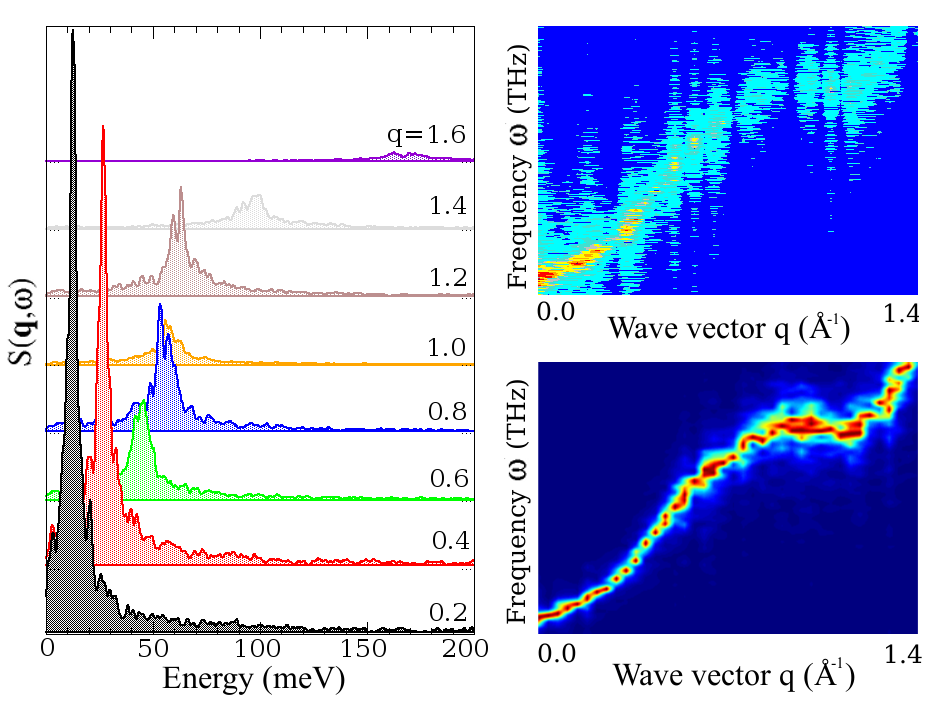}
  \caption{\label{fig:speelsasd}(Color online) Left panel: intensity of the dynamical structure factor $S^z(\mathbf{q},\omega)$ for a selection of $\mathbf{q}$-vectors along the crystallographic [001] direction for a ML Fe on W(110). Upper right panel: intensity plot of the ``raw" $S^z(\mathbf{q},\omega)$ directly obtained from an ASD simulation. Lower right panel: the same $S^z(\mathbf{q},\omega)$, obtained by following the peak-finding process described in the text. }
\end{figure}

\par
We obtain the spin wave dispersions by identifying the peak positions of the structure factor along particular directions in reciprocal space~\cite{Skubic2008,Tao2005,Chen1994}. This approach is based on the adiabatic approximation, and since longitudinal spin fluctuations are neglected, the interaction between spin waves and the Stoner continuum is not taken into account. However, Stoner excitations are usually present for short wave lengths, so the stiffness of the magnons should be captured by our method. Given the stochastic nature of the simulations, the calculated $S^k(\mathbf{q},\omega)$ spectra are very diffuse. Unless a considerable averaging of ensembles is performed, it can prove difficult to find the positions of the peaks corresponding to spin-wave excitations. In order to simplify the identification of the intensity peaks, we performed a post-processing scheme in which the intensity for each $\mathbf{q}$ vector is convoluted with a Gaussian, normalized to unity, and then used in a power function. This allows us to decrease the number of ensemble averages needed for each simulation by a few orders of magnitude.

\par
Fig.~\ref{fig:speelsasd} displays an example of the calculated dynamical structure factor $S^k(\mathbf{q},\omega)$, plotted after the convolution with a Gaussian function, for a selection of different $\mathbf{q}$ vectors along the [001] direction of the Fe/W(110) monolayer. For low $\mathbf{q}$ values, the spectral intensity of $S^k(\mathbf{q},\omega)$ is high, and decreases with increasing $\mathbf{q}$. At the same time, the peak width can be seen to increase with increasing energy. This is made clear by observing the peaks located at ${q}=0.8$ \AA$^{-1}$ and ${q}=1.2$ \AA$^{-1}$. These are positioned at similar energies, and do not exhibit noticeable broadening as a function of $\mathbf{q}$. The broadening of higher energy excitations can be explained by the damping term in the Landau-Lifshitz-Gilbert equation.

\par
In our proposed approach for calculating the magnon spectra, temperature effects enter in two ways, explicitly as the temperature at which the spin dynamics simulations are performed, and implicitly by the choice of which magnetic configuration the interatomic exchange interaction parameters are calculated from. In order to quantify temperature effects, we have calculated the spin wave stiffness of the Fe monolayer, calculated from LKGM simulations starting from both ferromagnetic and DLM states. The spin wave stiffness was evaluated by a least-squares fit of the magnon curves to a second order polynomial, within the range ${q}<0.2$ \AA $^{-1}$.  At $T=0$~K we calculated the spin-stiffness to be 160 meV \AA$^{2}$ when the $J_{ij}$s were obtained from a ferromagnetic configuration, and to be 110 meV \AA$^{2}$ from the DLM set of exchange interaction parameters. When introducing temperature to the spin dynamics simulations, we found that, for both the DLM and the ferromagnetic case, the spin wave stiffness decreased slowly up until the ordering temperature region, whereupon the softening became more pronounced. Around 120~K, the temperature at which the data of Prokop \emph{et al.}~\cite{Prokop2009} was measured, we find a non-negligible softening of the magnons of approximately 15\%, relative to 0~K, for each set of exchange interaction parameters. Thus, a correct description of the system lies somewhere in between the ferromagnetic and the DLM descriptions, which correspond to low- and high-temperature states, respectively.

\par
\begin{figure}
  \includegraphics[width=8.2cm]{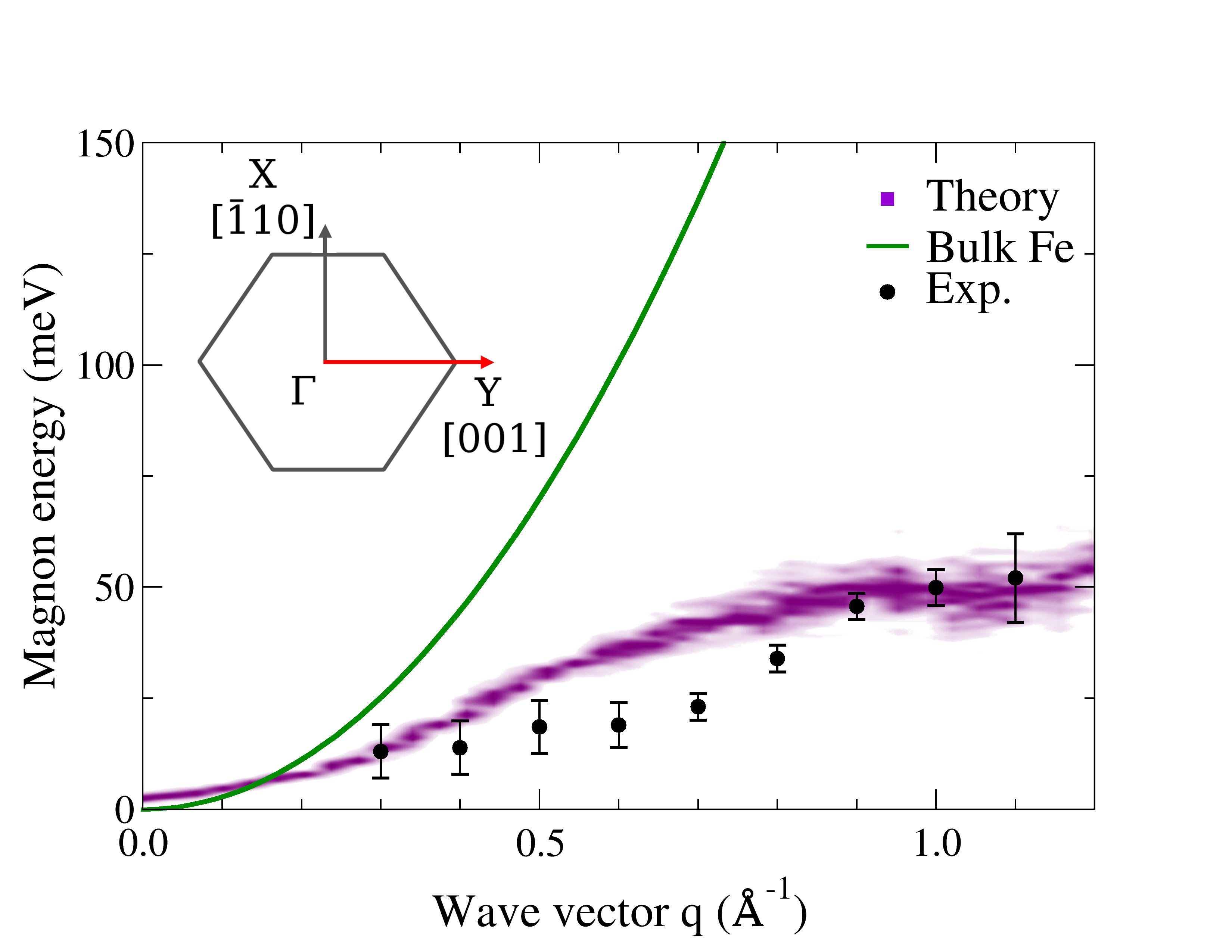}
  \caption{\label{fig:expspinwave}(Color online) Comparison between magnon dispersion curves along the [001] direction for a ML Fe on W(110). The dots are experimentally obtained data (Ref.~\onlinecite{Prokop2009}), whereas the thick purple line represents our numerically obtained data. For comparison, the experimental spin wave spectrum of bulk bcc Fe (corresponding to a spin wave stiffness constant of 280 meV \AA$^{2}$) (Ref.~\onlinecite{Mook1973}) is also displayed.}
\end{figure}

In order to make a direct comparison with the SPEELS data reported for the Fe/W(110) monolayer system~\cite{Prokop2009}, we therefore considered a partially disordered magnetic state with an average magnetization corresponding to 120~K. At this temperature, spin dynamics simulations resulted in an average magnetization of 85\% of the saturation magnetization. Consequently, we performed an electronic structure calculation for a partial DLM configuration corresponding to this state, and subsequently carried out an atomistic spin dynamics simulation based on the obtained exchange interaction parameters. The calculated magnon spectrum is displayed in Fig.~\ref{fig:expspinwave}, together with the experimental data of Prokop \emph{et al.}~\cite{Prokop2009}, and the experimentally determined spectrum for bulk bcc Fe~\cite{Mook1973}. The agreement with the SPEELS data is good, particularly for wave vectors around ${q}=1$ \AA$^{-1}$. Furthermore, the softening of the magnons in the monolayer system is captured. The estimate for the spin wave stiffness from our spin dynamics simulation at $T=120$~K with partial DLM exchange interaction parameters is 105 meV \AA$^{2}$, which compares well with the experimental fit of 74 meV \AA$^{2}$.

\par
In our calculations, relativistic effects have only been taken into account in the form of the magnetic anisotropy energy. It has recently been demonstrated that the Dzyaloshinskii-Moriya interaction is also important for the present system, most notably by giving rise to asymmetric magnon spectra around the $\Gamma$ point in the Brilluoin zone~\cite{Udvardi2009,Zakeri2010}. Although we have reproduced this result by inputting physically reasonable Dzyaloshinskii-Moriya vectors in the second term of Eq. (\ref{eqn:so}) (data not shown), we do not find that the DMI gives rise to any significant change in the overall agreement between theory and experiment shown in Fig.~\ref{fig:expspinwave}.

\par
In conclusion, we have shown that combining first principles calculations with atomistic spin dynamics simulations provides a powerful tool for studies of magnetic excitations in low dimensional systems. Applied to the reference system of a monolayer Fe on W(110), the method captures a significant magnon softening compared to bulk Fe, a finding which is in agreement with recent measurements. The microsopic reason why the magnon curve of this system is much softer than bulk bcc Fe can be found in three intertwined mechanisms. Firstly, the chemical relaxation between the Fe and W atoms influences the hybridization between Fe and W states, which modifies the exchange interactions. Secondly, the magnetic relaxation suggests that a partial DLM configuration is appropriate for the experimental situation at hand, and this configuration also reduces the exchange interaction strenghts. Finally, the dynamic effects also soften the magnon curve, which is captured by finite temperature simulations.

\begin{acknowledgments}
Financial support from the Swedish Research Council (VR), the European Research Council (ERC), the European Comission (EC), the Swedish Foundation for International Cooperation in Research and Higher Education (STINT) and the Carl Tryggers Foundation is acknowledged. Computer simulations were performed at the Swedish National Computer centre (NSC). We gratefully acknowledge L. Nordstr\"om, A. Delin, E. Papaioannou and R. B. Muniz for valuable discussions.
\end{acknowledgments}

\end{document}